\documentclass{IEEEtran}

\usepackage{cite}

\usepackage[pdftex]{graphicx}
\graphicspath{{./images/}}
\DeclareGraphicsExtensions{.jpg}

\usepackage{amsmath}

\usepackage{amssymb}

\usepackage{url}

\usepackage{lipsum}
\usepackage{dirtytalk}
\usepackage{subcaption}
\usepackage{amsfonts}
\usepackage{color}

\usepackage{siunitx}

\let\labelindent\relax
\usepackage[inline]{enumitem}

\usepackage{tikz}
\usetikzlibrary{calc,positioning,arrows}

\newcommand{\percent}[1]{$\approx$\num[round-mode=places,round-precision=4]{#1}\%}
\newcommand{\formatnum}[1]{\num[round-mode=places,round-precision=4]{#1}}
\newcommand{\approximate}[1]{$\approx$\num[round-mode=places,round-precision=4]{#1}}

\newcommand{\imageref}[4][0.5]{
  \begin{figure}
    \includegraphics[width=#1\textwidth]{#2}
    \caption{#3}
    \label{fig:#4}
  \end{figure}
}

{\bfseries}{\itshape}

\usepackage{algpseudocode,algorithm,algorithmicx}

\newcommand*\Let[2]{\State #1 $\gets$ #2}
\algrenewcommand\algorithmicrequire{\textbf{Precondition:}}
\algrenewcommand\algorithmicensure{\textbf{Postcondition:}}

\algnewcommand\algorithmicswitch{\textbf{switch}}
\algnewcommand\algorithmiccase{\textbf{case}}
\algnewcommand\algorithmicassert{\texttt{assert}}
\algnewcommand\Assert[1]{\State \algorithmicassert(#1)}%
\algdef{SE}[SWITCH]{Switch}{EndSwitch}[1]{\algorithmicswitch\ #1\ \algorithmicdo}{\algorithmicend\ \algorithmicswitch}%
\algdef{SE}[CASE]{Case}{EndCase}[1]{\algorithmiccase\ #1}{\algorithmicend\ \algorithmiccase}%
\algtext*{EndSwitch}%
\algtext*{EndCase}%

\begin{document}

\title{DHYMON: a Continuous Decentralized Hybrid Monitoring Architecture for MANETs}

\author{

  \IEEEauthorblockN{Jose Alvarez and Stephane Maag}
  \IEEEauthorblockA{
    SAMOVAR, Telecom SudParis, Universit\'e Paris-Saclay\\
    9 Rue Charles Fourier, 91000, Evry, FR\\
    \{Jose\_Alfredo.Alvarez\_Aldana,\\
    Stephane.Maag\}@telecom-sudparis.eu
  }

\and

  \IEEEauthorblockN{Fatiha Za\"{i}di}
  \IEEEauthorblockA{
    LRI-CNRS, Universit\'{e} Paris Sud, Universit\'e Paris-Saclay\\
    15 Rue Georges Clemenceau, 91400, Orsay, FR\\
    Fatiha.Zaidi@lri.fr
  }
}

\maketitle

\begin{abstract}
We introduce a novel decentralized monitoring algorithm for mobile ad-hoc networks. 
This algorithm is a combination of gossip-based and tree-based approaches. 
Its main feature is on multi root nodes selection which provides an opportunity to obtain more accurate results as it maximizes the coverage of the network. 
Our proposal relies on a thorough defined algorithm and its efficient and effective implementation. 
The algorithm is divided in two main parts that are its query procedure and its disseminate procedure.
To assess different parameters such as accuracy and convergence time, we conducted experiments with an in-house emulated test bed based on Docker and NS3.
To show the effectiveness and scalability of our proposal, we conduct intensive emulations that demonstrate very promising results regarding the gain of accuracy.
\end{abstract}

\section{Introduction} \label{introduction}

Mobile ad hoc networks (MANET) enable a collection of mobile hosts to communicate among themselves without any network infrastructure. 
Application domains have first arose for disaster situations or battlefields in order to deploy networks without base station of fixed networks. 
In the last decade, applications context range from health, home, vehicular, road safety and so on.
It has been widely studied in P2P, DTN or decentralized static networks with different approaches \cite{wuhib2009robust,kempe2003gossip,guerrieri2010distributed}.

Though MANET monitoring is crucial, it faces several issues mainly due to the inherent nature of such networks, i.e., the nodes mobility and the resources constraints. Some approaches have been proposed with a centralized node, named a coordinator, in charge of the monitoring. Unfortunately, such approaches suffer from many drawbacks due to energy efficiency, Internet access, infrastructure or other parameters, which render such approaches not always applicable.

Monitoring a network implies to obtain a global view of the system by means of attributes to be observed. In the literature, within a centralized approach, a central node is defined to collect and disseminate the observations. In \cite{cormode2013continuous}, the authors review the different communication mechanisms to optimize the data interchange. Such approaches are efficient for certain types of topology, whilst they are not at all in the presence of dynamic topology.
To overcome this limitation, decentralized approaches have gained considerable research interest.



The benchmarks provided by \cite{stingl2012benchmarking} established some non-functional requirements of a decentralized monitoring mechanism. 
Based on it, 
existing solutions rely on gossip-based or hierarchical-based approaches.
Gossip-based approaches demonstrate their robustness and stability in dynamic scenarios and changing topology.
Nonetheless, depending on the scalability, the cost and performance can be impacted.
On the other side, hierarchical approaches show an efficient performance, cost and scalability, although the robustness and stability may decrease in dynamic scenarios.
This shows that the two major categories perform very good under different characteristics, requirements and constraints of a network.
Therefore, we are convinced that a new algorithm could be derived from these two approaches for wider scenarios. The contributions of our paper are manifold.
\begin{itemize}

\item The introduction of a multi-root node approach enhancing the monitoring process and minimizing network fragmentation. 
By defining several root nodes, we considerably improve the accuracy results. 
\item We have conducted a large campaign of experimentation to assess the results of our proposal. For that purpose, we conducted emulations with Docker and NS3. To address the scalability issue, we ran a large range of experimentation which allows to be more confident in the obtained results.
\end{itemize}

In the remainder, 
the Section \ref{methods} explains our monitoring algorithm with its formalization. Section \ref{results} is devoted to the intensive experiments that have been conducted using NS3 and Docker through a configurable emulated testbed. 
We illustrate and discuss the effectiveness of our mechanisms. In Section \ref{relatedworks}, we present some interesting related works from which we got inspired
and finally, we conclude our paper by presenting conclusions and future works in Section \ref{conclusions} .


\section{Preliminaries} \label{preliminaries}

Network monitoring is an extensive field of interest.
It can be described as \say{a number of observers making observations and wish to work together to compute a function of the combination of all their observations}~\cite{cormode2013continuous}. 
The goal is that all observers (network nodes) compute a value $f(t)$ in a given instant of time $t$ in a collaborative way.
For our purposes, the function $t\mapsto f(t)$ [$\mathbb{R}^{+*}\rightarrow X$, $X$ being the domain targeted by $f$] is a linear and non-complex function like the average CPU 
or any other nominal value.

\subsection{Types of Monitoring}

The classification of the monitoring process has been studied in \cite{battat2014monitoring}.
For the purposes of this paper, we consider the two major types: centralized and decentralized.
The centralized type of monitoring, as stated by \cite{cormode2013continuous}, is when all the nodes report their observations to a central entity, named centralizer or coordinator.
The decentralized approach, deals with networks where there is no centralizer entity.
This implies that the network by itself needs to achieve a global view of a property of the network.
As stated by \cite{stingl2012benchmarking}, the more noticeable decentralized approaches are currently gossip and hierarchical.

\subsubsection{Gossip-based approaches}

The gossip approaches are based on the gossip or epidemic algorithms.
Gossip algorithms rely on selecting, from a set of reachable nodes, a random or a specific node (depending on the algorithm) to forward the data packet.
Epidemic algorithms try to forward the packet not only to one but to multiple nodes.
Flooding is the most common and simple algorithm for epidemic algorithms.
Gossip based monitoring algorithms have the advantage of being highly stable and perform better in increasing dynamic networks.
However, it may generate lots of traffic and, under certain scenarios, require more time to compute a value.

\subsubsection{Hierarchical-based approaches}

Hierarchical approaches commonly use tree structures in which the leaf nodes communicate the values with their parent node.
This is done recursively until it reaches the root node of the hierarchy.
These approaches consider a mechanism of either pulling data or pushing data with their nearby nodes.
To apply a hierarchical algorithm over a network, it is needed to build the topology before being able to monitor.
The advantages of hierarchical based monitoring algorithms are that they provide a fast convergence of the monitored property and produce less traffic.
The disadvantages of these solutions are that they are prone to errors in the event of a crash in the network. This means that it does not perform efficiently in a highly dynamic environment.

\subsection{Root node}

For our approach, we refer to every device in the network with communication capabilities as a node.
We consider every node to have the same set of features and no node has any outstanding property different to the other.
This being stated, each node can behave as a normal node or as a root node.
The root node plays an important role since it is the start of the monitoring process.
It will also be the node in charge of taking the start and end time of the process to compute the convergence time.
In the study \cite{battat2014monitoring}, mentions that one of the key challenges in MANET monitoring is survivability among others.
This challenges include heterogeneity, minimal human intervention, scalability, adaptability and many others.
To tackle this challenges, we define that every node in the MANET has the ability to act as the root node.
There are multiple studies in the literature (e.g. \cite{younis2008strategies}), that suggests that there is a point in the network where a node can optimize the area coverage.
Therefore for the purposes of this study, we randomly select the root nodes in an autonomous way.
A root node, will start its own monitoring process which will not affect any other monitoring process of any other root node.
Each process is running in the same network but isolated at the application layer to avoid clashes.

\section{A Decentralized Hybrid Monitoring Architecture} \label{methods}


In our past works~\cite{alvarez2016manets,AlvarezMZ17}, we studied a hybrid monitoring architecture where a random root node is selected to start and conduct the monitoring process.
These studies have shown promising results but there is a common challenge regardless of the number of nodes, size of the network and speed of the nodes, which is the network fragmentation.
This challenge directly affects the coverage of the algorithm. 
The coverage (defined in Section \ref{hybridArchitecture}) of the algorithm refers to the different observations collected by the monitoring process at the end by the root node.
We have observed in the preceding experiments and other works, that a dynamic network directly affects the network fragmentation due to mobility.
Fragmentation leads to partial observations given that the nodes in the network are not in reach all the time.
Assuming that all the nodes will always have a neighbor node is a strong assumption which can lead to over optimistic results from a non-realistic environment.
Therefore, we have opted to tackle this issue by introducing this concept and aiming to improve the overall monitoring process with more realistic conditions.

Aiming to maximize the coverage of the algorithm, we propose having multiple root nodes in order to cover more nodes who are left out due to the network fragmentation or any other communication problem due to mobility.
For the purposes of this study, we have randomly selected two root nodes in order to improve the monitoring efficiency analyzed in Section \ref{results}.

\subsection{Our Decentralized Hybrid Architecture} \label{hybridArchitecture}

Our proposal, namely DHYMON, defines an architecture based on a temporary tree structure supported by two mechanisms: the query and aggregate procedures. 
A root node is randomly selected and triggers the monitoring process.
This node will start by propagating a query and assembling a temporary tree (defined in \ref{vht}).
From this structure, it will aggregate the results until converging again in the root node.
Our approach is described in the Algorithm~\ref{alg:treesip_alg}.
The procedure is summoned when a packet (referred as payload) arrives on the input queue.
From this point, it decides what to be done based on the state of the node.

Depending on our algorithm described below, each node is capable of being in three possible states:\\
- Non-covered: meaning that a node was not reached by the query procedure then not included in the aggregate procedure. This could be derived from fragmentation or availability.\\
- Partially covered: meaning that a node was reached by the query procedure but it was not able to send and acknowledge the aggregate packet. This could happen due to mobility.\\
- Full covered: meaning that a node was reached by the query procedure. It was able to send the aggregate packet and it was acknowledged by its parent node.

The coverage of our algorithm is defined by the number of nodes fully covered by the monitoring process. 
The accuracy 
is the ratio between the coverage and the total number of nodes.
It provides a metric to measure how accurate is our algorithm.

We introduce the concept of monitoring based on multi-root nodes for a hybrid architecture. 
We aim to improve the overall accuracy by covering more nodes.
For this study we have selected two random root nodes.
Each root node triggers a separate and parallel monitoring process, each one of them taking into account its own coverage.
Then we analyze the coverage of the joint monitoring process to enhance the coverage and the accuracy of our algorithm over the network.
Making reference to our coverage definitions, this means that each fully covered node is taken into account. 
Then we will distinguish uniquely each fully covered node in any of the two root nodes.
Due to this process, there are important premises worth mentioning:
\begin{itemize}
	\item The overall traffic generation from our proposal will increase probably by a factor of two.
	\item Ideally, where all nodes are fully covered by both root nodes, there will be a coverage overlapping, which will not lead to any improvements. 
	\item In the worst case scenario, the gain 
	will be minimal, maybe including a couple of nodes to the overall result leading to a small gain in coverage and accuracy.
	\item In the best case scenario, the nodes will be able to cover more of the network. 
	This will result in high coverage and accuracy gains. We believe that this is the motivating scenario, which will demonstrate the efficiency of our approach. This scenario could be more prominent in dense and dynamic networks.
\end{itemize}

\begin{algorithm}

\tiny 
  \caption{Procedure that attends the incoming messages}
  \label{alg:treesip_alg}
  \begin{algorithmic}[1]
    \Require{$state$ is defined and initialized as $INITIAL$, the $SEND*()$ functions trigger automatically a timer, the $sendAggregate()$ function aggregates internally results and observations}
    \Statex
    \Procedure{attendBufferChannel}{$payload$}

	\Switch{$state$}
		\Case{$INITIAL$} \label{algo:initial_state} 
			\If{$payload.Type = START$ $\|$ $QUERY$} 

				\Let{$state$}{$Q1$}
				\State \Call{sendQuery}{payload} 
			\EndIf

		\EndCase
		
		\Case{$Q1$} \label{algo:q1_state}
			\If{$payload.Type = QUERYACK$}
				\Let{$state$}{$Q2$}
				\Let{$queryACKList$}{append $payload.Source$}
			\ElsIf{$payload.Type = TIMEOUT$} \label{algo:q1_state_timeout}
				\Let{$state$}{$A1$}
				\State \Call{sendAggregate}{parentIP, result}
			\EndIf
		\EndCase
		
		\Case{$Q2$} \label{algo:q2_state}
			\If{$payload.Type = QUERYACK$}
				\Let{$state$}{$Q2$}
				\Let{$queryACKList$}{append $payload.Source$}
			\ElsIf{$payload.Type = AGGREGATE$}
				\Let{$state$}{$A1$}
				\Let{$queryACKList$}{remove $payload.Source$}
				
					\If{\Call{empty}{queryACKList}}
						\State \Call{sendAggregate}{parentIP, results}
					\EndIf
			\EndIf
		\EndCase
	
		\Case{$A1$} \label{algo:a1_state}
			\If{$payload.Type = AGGREGATE$}
				\Let{$state$}{$A1$}
				\Let{$queryACKList$}{remove $payload.Source$}
				
					\If{\Call{empty}{queryACKList}}
						\State \Call{sendAggregate}{parentIP, result, observations} \label{algo:a1_state_sendag}
						
						
					\EndIf
			\ElsIf{$payload.Type = TIMEOUT$} \label{algo:a1_state_timeout}
				\Let{$state$}{$A2$}
				\State \Call{sendRoute}{payload}
			\ElsIf{$payload.Type = AGGREGATEACK$} \label{algo:a1_state_done}
				\Let{$state$}{$INITIAL$} 
				\State \Call{done}{}	
			\EndIf
		\EndCase
		
		\Case{$A2$} \label{algo:a2_state}
			\If{$payload.Type = AGGREGATEACK$} \label{algo:a2_state_done}
				\Let{$state$}{$INITIAL$} 
				\State \Call{done}{}	
			\ElsIf{$payload.Type = TIMEOUT$} \label{algo:a2_state_timeout}
				\Let{$state$}{$A3$}
				\State \Call{sendForward}{payload}
			\EndIf
		\EndCase

		\Case{$A3$} \label{algo:a3_state}
			\If{$payload.Type = AGGREGATEACK$} \label{algo:a3_state_done}
				\Let{$state$}{$INITIAL$} 
				\State \Call{done}{}	
			\ElsIf{$payload.Type = TIMEOUT$ $\&$ !\Call{empty}{relayList}} \label{algo:a3_state_timeout}
				\Let{$state$}{$A2$}
				\Let{$relay$}{$pop relaySetList$}
				\State \Call{sendForward}{payload}
			\ElsIf{$payload.Type = TIMEOUT$ $\&$ \Call{empty}{relayList}} 
				\Let{$state$}{$INITIAL$}
				\State \Call{error}{}	
			\EndIf
		\EndCase
	\EndSwitch
    \EndProcedure
  \end{algorithmic}

\end{algorithm}

\subsection{Virtual Hierarchical Topology} \label{vht}

In our approach, a virtual hierarchical topology (VHT) for a time window is built to process a hierarchical based
aggregation of the monitored values through the network. 
The VHT concept has been introduced in \cite{huang2005virgo} and adapted by Google in one of his patents for cache nodes \cite{Goooglepatent}. 
The advantage of the VHT is its simple packet forwarding configuration. 
Each child node only forwards data packets to its parent node. 
The message will propagate in such a manner until it reaches the root node. 
The following steps summarize the VHT construction:
\begin{itemize}
	\item Each source node sends a query to its neighbors. A timestamp information labels the time window.
	\item Each node (that is not an edge node) receiving a query forwards it if not already received before. The hierarchy father/child and the timestamp are stored.
	\item An edge node receiving a query does not forward it.
\end{itemize}

This concept is very important since we here rely on a hierarchical approach.
Therefore, we need a topology built before we are able to aggregate our results.
But our goal is to utilize this hierarchical topology only for a limited and minimal amount of time so we minimize its frangibility.
For that, we use the VHT only during a limited time during which the monitoring is done.
It is a virtual representation since, given the mobility, it will probably not be able to keep the same topology physically.
The VHT validity time starts with the monitoring process and ends with the convergence of the process, both instances of time considered from the root node.
This way, we can aggregate efficiently and for every new monitoring process, a new VHT is derived.



\subsection{Query Procedure} \label{queryProcedure}

The query procedure refers to the process of propagating in an epidemic way the monitoring packet.
This query is forwarded in an epidemic approach to the nodes to optimize the time to disseminate the corresponding information to all the network.
This is performed recursively until the edge of the network is reached.
This process creates the VHT.

It is worth mentioning that based on the work of Drabkin et. al.~\cite{drabkin2007rapid}, it is expected that the query takes from 2 to 5 hops to reach the majority of the network.
This is calculated based on the coverage probability of a message $m$ reaching all the nodes in the network.
For our case in particular, it means that theoretically, the VHT will have from 2 to 5 levels.

\subsection{Aggregate Procedure} \label{aggregateProcedure}

Once the data is disseminated up to the edge of the network, the edge nodes change to aggregate procedure and start sending recursively their results of the query to their parent up to the root node.
This process is an aggregation of all the data of a node and its children in order to collect the monitored values.
We define the transition between the query procedure to the aggregate procedure by using a timeout mechanism.
Depending on the mobility properties of the network, this timeout can be sufficient to break some physical visibility between some parents and child nodes.
We introduce a gossip routing fallback to deliver the package to the corresponding node. 

\subsection{Gossip Routing Fallback} \label{gossipRoutingFallback}

\begin{figure}
\centering

\scalebox{0.70}{%
\begin{tikzpicture}[->,>=stealth',shorten >=1pt,auto,node distance=5cm,
  thick,main node/.style={circle,fill=blue!20,draw,
  font=\sffamily\small\bfseries,minimum size=8mm}]

  \node[main node] (I) {Initial};
  \node[main node] (WR) 	[right of=I] {WR};
  \node[main node] (DN) [right of=WR] {DN};

  \path[every node/.style={font=\sffamily\small,
      fill=white,inner sep=1pt}]

  (I) edge [] node[align=center] {$(Aggregate||Route)$/$Hello$} (WR)

  (WR) edge [] node[align=center] {$HelloReply$/$Route$} (DN)
  
  (I) edge [loop below] node[align=center] {$Hello$/$Reply$} (I)

  (WR) edge [loop below] node[align=center] {$Timeout$/$Hello$} (WR)
  
  (DN) edge [loop below] node[align=center] {$Reply$/$ReplyAck$} (DN)
  
  (DN) edge [bend right=25] node[align=center] [above] {$Hello$/$HelloReply$} (I);

\end{tikzpicture}
}

\caption{IO FSM definition of the fallback gossip routing process for the aggregate procedure} \label{fig:automata}
\end{figure}
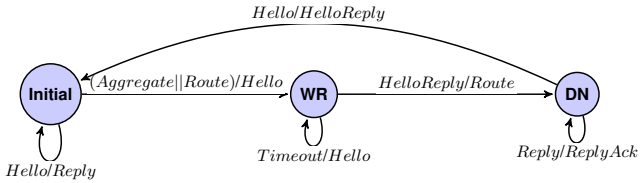

We formalize this fallback process as the input/output finite state machine (FSM) depicted in Figure \ref{fig:automata}.
This is an important part of the architecture since due to the mobility, it is expected to support the VHT.
Every packet is uniquely identified by the concatenation of the unique identifier of the generating node (IP address) and the timestamp of the creation time (unix time).
The set of states is $S=\{Initial, WR, DN\}$, $WR$ refers to waiting for a message hello reply and $DN$ refers to done routing the message.
We define the set of inputs as $I=\{Aggregate, Route, Hello, HelloReply\}$ and the set of outputs as $O=\{Hello, HelloReply, Route\}$.
Then we define the transition as $Timeout$, which is any $t$ that maximizes the delivery of the packet.
The $Aggregate$ message triggers the routing process from the monitoring layer.
The process aims at sending a $Hello$ message to its neighbors, the closest node will reply with a $HelloReply$.
Then it will route the packet to this node, theoretically the state $DN$ should be the final state of the FSM.
But there is a probability that, due to mobility, the packet needs to be rerouted again by it, therefore we leave the possibility open by the transition between $DN$ and the initial state.
If we received a $Route$ message, we will first verify if it is addressed to itself.
If this is the case, we will route the packet internally to the monitoring layer, otherwise we will continue the routing process.

\section{Experiments} \label {results}


We evaluate our proposal using an emulator\footnote{https://github.com/chepeftw/NS3DockerEmulator} built in-house based on Docker (v17.03.1-ce) and NS3 (v3.26).
It sits on top of an orchestrator that relies on Amazon EC2 to launch parallel emulations for fast and efficient generation of results.
Finally, the orchestrator parses the logs from the Docker containers per emulation, extracts the information and centralizes it in a MongoDB server for further analysis.
The testbed consisted in an implementation\footnote{https://github.com/chepeftw/Treesip} of our proposal in the language Go (v1.8).
The idea was to identify primarily the convergence time and the accuracy.
The convergence time is described as the time it takes from the moment that the monitoring started by the root node, to the moment that the root node was able to return a verdict.


We define two different suites of tests.
First, we analyze the timeout selection. 
In past works, a random timeout of 1800ms was used and we concluded that this value affects directly the convergence.
Therefore, we simulate the single root node varying the timeout using different number of nodes.
This allows to find a pattern for a better timeout maximizing the convergence time.
Second, we study the convergence and accuracy of the multi-root node approach. 
For this case, we simulate multiple number of nodes using different network sizes varying the speed of the nodes and finally comparing this to the single root node results varying the same parameters.

 
For all scenarios, the MAC protocol is 802.11a, the transport protocol is UDP with a data rate of 54Mbps.
Each node had a range of $\approx$125m. 
We rely on the random waypoint mobility model, using a pause time equal to 0 providing therefore continuous mobility.
All scenarios were intended to test the convergence time and accuracy in a mobile environment. 
All simulations have an initial configuration time to randomize the position of the nodes. 
The idea is for the nodes to shuffle from their original location, regardless of the fact that they are located with a random pattern by NS3.
All of this is done to enhance the trustability of the results by using random values to ensure that the algorithm works in any given scenario and that it is not tied by any emulation parameter.
Given the node range, we decided that every 100m$\times$100m of network size should have at least 2 nodes, then based on this requirement, we calculated the network size and the number of nodes.
The parameter relationship between number of nodes, network size and Amazon EC2 instance type is summarized in Table~\ref{table:nodenumberrelations}.
Then, for each separate arrangement of parameters, the emulations ran for 100 cycles to ensure stable and consistent results.

\begin{table}
  \caption{Emulation network parameters}\label{table:nodenumberrelations}
  \vspace{-1em}
  \begin{center}
      \begin{tabular}{| c | c | c |}
      \hline
      \textbf{Nodes} & \textbf{Network Size (m)} & \textbf{EC2 type} \\ \hline
      20 & 316x316 & t2.small  \\ \hline
      30 & 387x387 & t2.small  \\ \hline
      40 & 447x447 & t2.medium  \\ \hline
      50 & 500x500 & c4.large  \\ \hline
      60 & 548x548 & c4.large  \\ \hline
      \end{tabular}
  \end{center}
  \vspace{-2em}
\end{table}


\subsection{Timeout Selection Experiment} \label{timeoutselectionexperiment}

The results of this experiment for different number of nodes moving at a speed of 2m/s and 10m/s can be seen respectively in Fig. \ref{fig:timeoutResults2ms} and Fig. \ref{fig:timeoutResults10ms}.
Each data point in this graph represents 100 emulation runs using the same parameters, this means that each graph represents 3,000 emulation cycles.

\imageref[0.47]{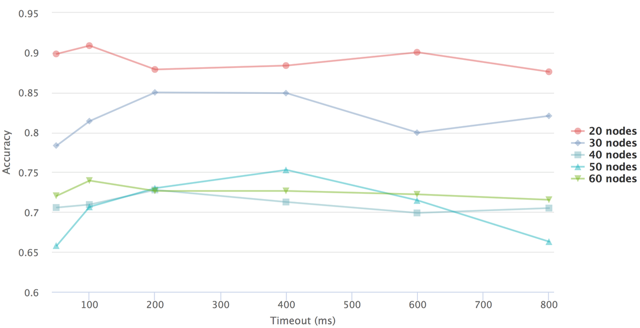}{Timeout vs accuracy for nodes at 2m/s}{timeoutResults2ms}
\imageref[0.47]{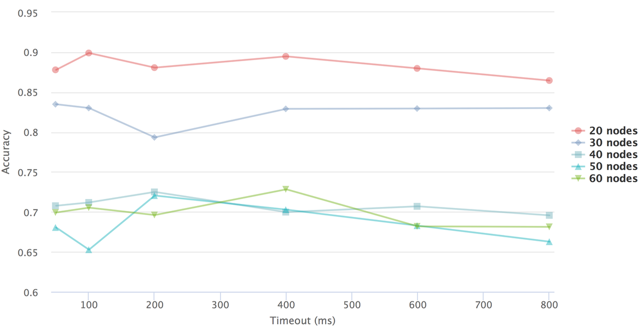}{Timeout vs accuracy for nodes at 10m/s}{timeoutResults10ms}

The results in Fig. \ref{fig:timeoutResults2ms} show that the most prominent and promising timeouts are 200 and 400.
But in Fig. \ref{fig:timeoutResults10ms}, it seems that 50 and 200 could be the most prominent and promising in general.
We may conclude from these graphs that there is no pattern and that the underlying problematic could be more complex that we thought. 
If we analyze this experiment from another perspective (Table \ref{table:timeoutresultsgroupedbytimeout}), we can group the results by timeout from the different node speeds and different number of nodes sort by accuracy.
Each row of this table represents 2,500 cycles of emulation.
Also the convergence column shows the approximate average value, followed by the minimum and maximum value between the parenthesis.
We can observe three interesting results.
The empirical assumption would be that the timeout and the accuracy could have a direct correlation, but from the data we can see that this is not the case.
Based on our past studies speculations, it shows a direct correlation between the timeout and the convergence time.
We can observe different average convergence times, but if we take a closer look at the minimum value of the samples, it shows that in the best case scenario, the process would take almost the time of the selected timeout.
And the last remark is that, depending on the mobility, it seems that it can last 20 seconds to try to converge.
This shows that the worst case scenarios, it will take a lot of time due to the gossip approach but it will converge nonetheless.

\begin{table}
  \caption{Accuracy and convergence results by timeout}\label{table:timeoutresultsgroupedbytimeout}
  \vspace{-1em}
  \begin{center}
      \begin{tabular}{| c | c | c |}
      \hline
      \textbf{Timeout (ms)} & \textbf{Accuracy} & \textbf{Convergence (ms)} \\ \hline
      50 & \approximate{0.769455337690633} & \formatnum{2523.66063348416} (\formatnum{55}, \formatnum{16253})  \\ \hline
      400 & \approximate{0.764181937172778} & \formatnum{3080.22993019197} (\formatnum{403}, \formatnum{22564})  \\ \hline
      200 & \approximate{0.763175962523791} & \formatnum{2952.15239350022} (\formatnum{202}, \formatnum{21262})  \\ \hline
      600 & \approximate{0.761999707858605} & \formatnum{3503.95179666959} (\formatnum{603}, \formatnum{17372})  \\ \hline
      100 & \approximate{0.759603663003664} & \formatnum{2766.03296703297} (\formatnum{102}, \formatnum{18850})  \\ \hline
      800 & \approximate{0.755914589709956} & \formatnum{3920.74814167031} (\formatnum{804}, \formatnum{20477})  \\ \hline
      \end{tabular}
  \end{center}
  \vspace{-2em}
\end{table}

\subsection{Multi-Root Node Experiment} \label{multirootnodeexperiment}

\begin{figure*}
  \begin{subfigure}{.32\textwidth}
    \centering
    \includegraphics[width=\linewidth]{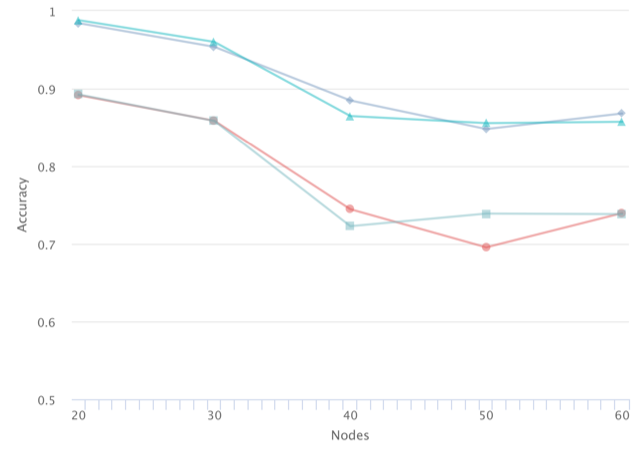}
    \caption{Node speed 2m/s}
    \label{fig:multirootnodeexperiments2}
  \end{subfigure}%
  \begin{subfigure}{.32\textwidth}
    \centering
    \includegraphics[width=\linewidth]{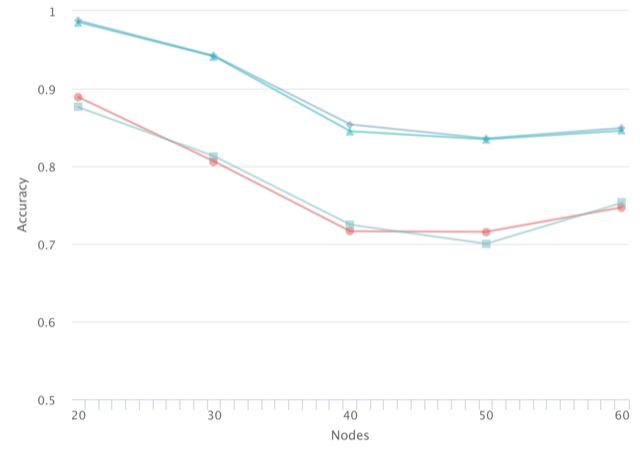}
    \caption{Node speed 6m/s}
    \label{fig:multirootnodeexperiments6}
  \end{subfigure}%
  \begin{subfigure}{.32\textwidth}
    \centering
    \includegraphics[width=\linewidth]{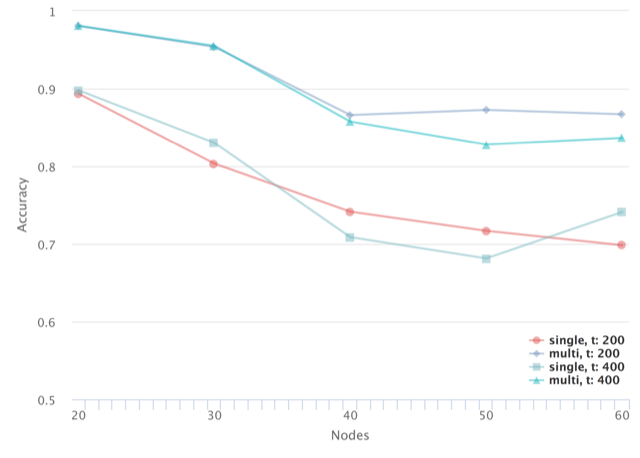}
    \caption{Node speed 10m/s}
    \label{fig:multirootnodeexperiments10}
  \end{subfigure}%
  

%

  \caption{Accuracy vs number of nodes with single-root node and multi-root nodes for different timeouts and different speed.}
  \label{fig:multirootnodeexperiments}
\end{figure*}

From our preceding studies, we have seen that the general accuracy of the monitoring process is approximately \percent{80}.
In this study, we intend to enhance the accuracy by introducing a second root node.
As defined in \ref{hybridArchitecture}, the coverage is based on the nodes which were able to be fully covered.
It means that different monitoring processes performed by two different root nodes could have a different set of fully covered nodes.
This difference could be none, meaning that the two processes were identical which is the theoretical best case scenario, but not realistic.
The difference could be more than one node and then could lead to a major enhancement.
To calculate these sets and differences, we analyzed through the logs which nodes were fully covered by which monitoring process.
This provided two distinct coverage lists that we crosschecked. 
By doing this, we concluded on a list with the nodes covered by the two processes.
The results are summarized in Figure \ref{fig:multirootnodeexperiments}.
Every data point of every graph represents 100 emulation cycles, therefore every graph contains 2,000 emulation cycles.

To show the improvements of our proposal, we ran again, under the same conditions, the emulations for single-root node approach and we illustrate them in the same graph as their multi-root node approach counterpart.
We did this comparison by using two different timeouts and different node speeds to ensure that the enhancements are consistent.
In the graphs, value $t$ refers to the timeout.

In all the graphs, we can note that there is a consistent enhancement between the single-root node and the multi-root nodes approach.
For this, we calculated the difference between all the corresponding data points and calculated that the average enhancement is \percent{12.873563579477612}.
The general accuracy for all the multi-root nodes experiments is \percent{90.0593510681179}.
From these results, it is fair to say that the multi-root nodes approach increases the general accuracy by \percent{12} (based on 10,000 cycles).
This is an interesting result since shows a solution to network fragmentation by covering the network from different nodes at the same time in a monitoring scenario.
This also opens the question about how many root nodes would be the ideal number for providing efficient monitoring and minimizing the network traffic.

To analyze the convergence time of the monitoring process for the 200ms timeout, we can refer to Table \ref{table:multirootnodeconvergence200}. 
Each row in the table represents 500 emulation cycles.
Also the convergence and tree depth columns show the approximate average value, followed by the minimum and maximum value between the parenthesis.
We can note that there is a direct correlation between the number of nodes and the time it takes to converge.
This is an expected behavior 
and it is also reflected in Section \ref{timeoutselectionexperiment}.
The overall average using a timeout of 200ms is \approximate{3066.98381222177}. 
We can observe 
that the minimum convergence time is determined by the timeout.
These results give a great view of the performance of our algorithm, suggesting it can converge in a relatively fast time while providing a great accuracy.
Regarding the tree depth, this confirms the hypothesis from 
Section \ref{queryProcedure}.
Therefore, we can state that the approach has an approximate of 4 to 6 levels of depth in the VHT.
Note that there are scenarios where the VHT can reach up to 16 levels and nonetheless it will be able to converge.
And in the best case scenario, it can rely on only 2 levels, which theoretically are more than enough to cover the majority of the network.
It is important to note that there are no scenarios with 1 level. This proves that the initial configuration time is important since it provides accurate scenarios.


\begin{table}
  \caption{Multi-root convergence results for 200ms timeout}\label{table:multirootnodeconvergence200}
  \vspace{-1em}
  \begin{center}
      \begin{tabular}{| c | c | c |}
      \hline
      \textbf{Nodes} & \textbf{Convergence (ms)} & \textbf{Tree depth} \\ \hline
      20 & \approximate{1307.37242798354} (212, \formatnum{20361}) & \approximate{4.03292181069959} (2, 8)  \\ \hline
      30 & \approximate{1856.72} (223, \formatnum{18907}) & \approximate{5.098} (2, 11)  \\ \hline
      40 & \approximate{3685.186} (220, \formatnum{16707}) & \approximate{5.158} (3, 9)  \\ \hline
      50 & \approximate{3657.32661290323} (206, \formatnum{13687}) & \approximate{5.88104838709677} (4, 16)  \\ \hline
      60 & \approximate{4822.38650306748} (202, \formatnum{15583}) & \approximate{5.92638036809816} (4, 11)  \\ \hline
      \end{tabular}
  \end{center}
  \vspace{-2em}
\end{table}


\subsection{Routing Layer Results}

From the multi-root nodes experiments, we analyzed how often was the routing layer in use and how it varied depending on the different number of nodes.
We analyzed the results by node speed as well but interestingly the results were stable. Meanwhile, with different number of nodes, there was a big variance.
The results are shown in Table \ref{table:routinglayerusage}.
Each row represents 500 emulation cycles.
It can be observed that the more the number of nodes, the more the messages are routed through the fallback mechanism.
This is congruent with the purpose of the routing layer, given that it reacts as a fallback mechanism to make the communication possible between a parent node and a child node in the VHT.
Our algorithm takes into account the fact that the more the nodes, the more the network fragmentation will occur.
It is also important to notice how the delivery ratio drops as the number of nodes increases.
These are interesting results since they point out the importance of this fallback mechanism and a feasible point for enhancements.

\begin{table}
  \caption{Routing layer usage based on the number of nodes}\label{table:routinglayerusage}
  \vspace{-1em}
  \begin{center}
      \begin{tabular}{| c | c | c |}
      \hline
      \textbf{Nodes} & \textbf{Messages Sent} & \textbf{Messages Received} \\ \hline
      20 & \approximate{8.93634496919918} & \approximate{8.21560574948665}  \\ \hline
      30 & \approximate{15.746} & \approximate{13.916}  \\ \hline
      40 & \approximate{35.7} & \approximate{29.81}  \\ \hline
      50 & \approximate{40.5754527162978} & \approximate{33.5513078470825}  \\ \hline
      60 & \approximate{48.6332665330661} & \approximate{39.7074148296593}  \\ \hline
      \end{tabular}
  \end{center}
  \vspace{-2em}
\end{table}

\section{Related Works} \label{relatedworks}

%
%

In the gossip based categorization, we can discuss about Gossipico \cite{van2012gossip}.
This is an algorithm to calculate the average, the sum or the count of node values in a large dynamic network.
The combination of two mechanisms, count and beacon, provides the networks nodes counting in an efficient and quick way.
This algorithm was tested in static networks. 

Relying on Gossipico, Mobi-G \cite{stingl2014mobi} for ad-hoc networks was proposed.
It is designed for urban outdoor areas with a focus on pedestrian that moves around by foot.
The idea is to create the global view of an attribute, which is built ideally incorporating all the nodes in the network.
This global view is disseminated to all the nodes in the network to inform the current system state.
It can provide accurate results even for fluctuating attributes. It also can reduce the communication cost.
It does not suffer from long range connectivity but the accuracy decreases for an increasing spatial network size.

On the hierarchical categorization, BlockTree \cite{stingl2013blocktree} proposes a fully decentralized location-aware monitoring mechanism for MANETs.
The idea is to divide the network in proximity-based clusters, which are arranged hierarchical.
Each cluster or block aggregates the data respecting a property.
The algorithm requires that all nodes from the same cluster or block are reachable within one hop.
Even though the good performance, the average power consumption increases directly proportional to the spatial network size or node density.

In \cite{RicherzhagenSRM15}, the authors present the adaptative monitoring mechanism CRATER. The architecture exploits the connectivity and resource characteristics of the mobile nodes to ease the continuous network monitoring by a dynamic reconfiguration of the monitoring topology. However, a strong assumption is that 
the MANET always needs a central server that has continuously a link with all nodes.

We should also cite \cite{stingl2012benchmarking} proposing a benchmark between both categories. 
They provide different workloads for the tests: baseline (idealized conditions), churn, massive join or crash, increasing number of attributes and increasing number of peers.
Under ideal conditions, the hierarchical approach outperforms the gossip approach.
In the presence of sudden topology changes, the gossip approach performs well and is able to continue working robustly. 


\section{Conclusions} \label{conclusions}

We have presented our DHYMON architecture based on a hybrid algorithm for monitoring decentralized networks which is a combination of gossip-based and hierarchical-based algorithms. The algorithm has been thoroughly defined which helps to be more confident in the implementation results.
The algorithm is constructed on top of two major procedures, the query and the aggregate.
The gossip-based approach is applied to the query procedure to disseminate the query in an efficient way.
Once the query is propagated, the network changes to the aggregate procedure. 
Besides, with the help of a time-based hierarchical approach, the computation of a global property is achieved. This computation is enhanced with a  multi root node selection which permits to enhance the network coverage.
We designed a scalable and configurable testbed using NS3 and Docker that illustrates the effectiveness of our approach for different scenarios.
An immediate line of future work is to introduce a new mechanisms of root nodes selection, for instance, it could be based on location, energy, computing power and other parameter.  



Another line of work will be to monitor complex functions such as the interoperability within a MANET.
For that purpose, we need to define an optimal solution to propagate a more complex function through our query mechanism.
This requires to analyze multiple interoperability approaches to provide a proposal for monitoring this process.
To monitor interoperability, we guess that it is needed to analyze not all observation points, 
but more specifically a subset of observation points.

\bibliographystyle{plain}
\bibliography{ms}


\end{document}